\documentstyle[aps,epsfig]{revtex}
\draft
\title{Nodal Domain Distribution for a Nonintegrable Two-Dimensional Anharmonic Oscillator} 
\author{Hirokazu Aiba$^1$~and ~Toru Suzuki$^2$}
\address{$^1$Kyoto Koka Women's College, 38 Kadono-cho Nishikyogoku, Ukyo-ku,
         615-0882 Kyoto, Japan\protect\\
     $^2$Department of Physics, Tokyo Metropolitan University, 
     192-0397 Hachioji, Japan}
\begin{document}
\twocolumn
\vspace{1cm}
\date{\today}
\maketitle
\begin{abstract}
We investigate the transition from integrable to chaotic dynamics in the quantum mechanical 
wave functions from the point of view of the nodal structure  by
employing a two dimensional quartic oscillator.
We find that the number of nodal domains is drastically reduced as the dynamics of the system 
changes from integrable to nonintegrable, and then gradually increases as the system becomes
chaotic. The number of nodal intersections with the classical boundary as a function of the 
level number shows a characteristic dependence on the dynamics of the system, too. 
We also calculate the area distribution of nodal domains and study the emergence of the power 
law behavior with the Fisher exponent in the chaotic limit. 

\end{abstract}
\pacs{PACS number: 05.45.Mt, 64.60.Ak}
\def\be{\begin{equation}}
\def\ee{\end{equation}}
\section{Introduction}

Quantum signatures of classically chaotic systems  have been intensively studied 
and are now known to have some universal features in the energy level statistics. 
Similar investigations on the signatures in the wave functions which may 
distinguish chaotic systems from integrable ones have also been 
performed\cite{Berry,McDonald}.  A related signature of  chaotic systems 
is given by the amplitude distribution of wave functions which empirically 
reproduces the results of random matrix theory\ \cite{Brody}.

Recently, it was suggested in Ref.\ \cite{Blum} that there is in fact such a universal 
character in the statistics of nodal domains of wave functions. The authors of
Ref.\ \cite{Blum} 
calculated the number of nodal domains $N_i$ of the $i$th wave function, i.e., 
the regions where the wave function 
has a definite sign without crossing zeros, for two-dimensional billiards. They 
showed that the distribution of normalized number $N_i/i$ of nodal domains for 
separable 
systems has a universal feature characterized by a square root singularity, while 
that for chaotic billiards shows a completely different behavior, suggesting a 
scaling law $N_i \approx i$ for large $i$. The latter scaling law has been 
derived in Ref.\ \cite{Bogomolny}, where the authors adopted a percolation-like model to 
count the number of nodal domains. 
The authors of Ref.\cite{Bogomolny} analytically derived the scaling law for the average 
number of nodal domains 
and showed that it agrees well with the numerical results for the superposition of random 
waves, i.e., a model which is supposed to simulate the wave functions for chaotic 
billiards\cite{Berry}. The percolation-like model allowed them to predict a power law 
behavior for the distribution of nodal domain areas, 
and also the fractal dimension of nodal domains. These predictions were shown to 
agree well with the numerical results for the superposition of random waves, although one 
may not conclude from these results alone that the area distribution  provides a 
clear signature of quantum chaoticity. 
The number of nodal domains was studied also experimentally for 
the chaotic microwave billiard\ \cite{Savytskyy}.

In the above studies of nodal domains, the two extremes of 
dynamical systems, completely integrable (separable) and chaotic (or its 
alternative), have been considered. It is the purpose of the present paper 
to extend these studies to a more general nonintegrable system, where, by 
controlling a parameter in the Hamiltonian, one can interpolate the two 
extremes. We expect this would show the transition from integrable to 
chaotic systems for the nodal domain distribution and thus may provide a clue 
to the role of nonintegrable perturbation which was implicit in the 
percolation-like model. 
We expect that a study of the power law behavior of 
nodal domain areas  would also  give a suggestion on 
the validity of the assumption adopted in the model.

Below we first describe the model and the numerical procedure. We present 
numerical results for the distribution of nodal domain numbers in Sec.\ \ref{Sec_domain} 
together with some analytical considerations. The distribution of
the nodal intersections with the boundary of the classically allowed region 
is presented in Sec.\ \ref{Sec_isect}. Results for the distribution of 
nodal domain areas are given in Sec.\ \ref{Sec_ns}. Proofs of formulas and some 
details of the calculations in the text are given in the Appendices.

\section{Model and Numerical Procedure}
\label{Sec_model}

\begin{figure}[tb]
\begin{center}
\epsfig{file=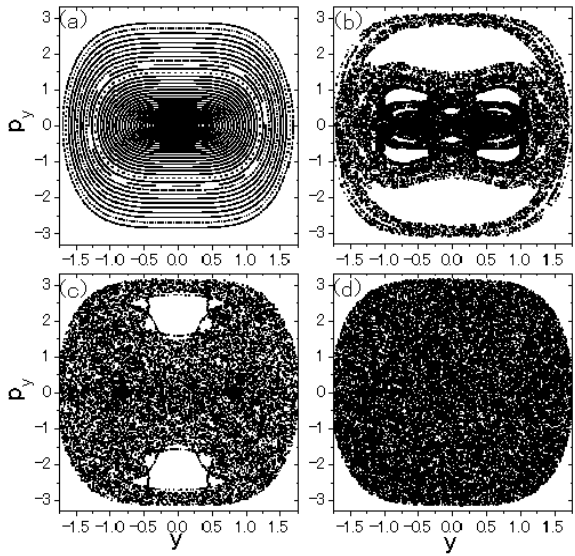,width=8cm}
\end{center}
\caption{Poincar\`{e} surface of section  for the system
 of the Hamiltonian Eq.\ (\ref{hamiltonian})
with $\alpha=1.01$ at  $E=5.0$, $x=0.0$, and 
(a)
$k=0.0$, (b) 0.2, (c) 0.4,  (d) 0.6. 
The abscissa axis represents $y$ coordinate and the ordinate $p_y$.
}
\label{Pmap_fig}
\end{figure}

As a model which incorporates integrable as well as almost 
chaotic systems, we adopt a two-dimensional quartic oscillator
\begin{eqnarray}
         H=\frac12(p_x^2+\alpha p_y^2)+V(x,y), \nonumber \\ 
         V(x,y)=\frac12(x^4+y^4)-kx^2y^2,
\label{hamiltonian}
\end{eqnarray}
where the parameter $k$ controls the dynamics of the system.
Detailed studies performed at $\alpha=1$ shows that 
the classical dynamics of the system at $k=0.0$ is integrable, 
becomes irregular as the value of $k$ increases, and reaches an almost 
chaotic system at $k=0.6$\cite{Mayer}. The energy level statistics of the 
quantum mechanical system show a similar transition, e.g., from Poisson to 
Wigner level spacing distribution as $k$ increases\cite{levels}.
The parameter $\alpha$ is introduced to break the symmetry with respect 
to the exchange of the $x$ and $y$ coordinates, which otherwise 
leads to an ambiguity in the definition of the eigenfunction at $k=0.0$. 
The value of $\alpha$ is set to the value 1.01 throughout.
We plot in  Fig.\ \ref{Pmap_fig} the Poincar\`{e} surface of section for the system 
with $\alpha=1.01$ at several values of $k$. Qualitative behavior is
almost the same as that with $\alpha=1$.
\vspace{-0.5cm}

In the numerical calculation, we first diagonalize the Hamiltonian 
in a large harmonic oscillator (h.o.) basis and obtain wave functions for 
eigenstates. To determine nodal domains we cut the two-dimensional sheet into 
small squares (meshes) so that 
they cover the allowed region of classical motion for each eigenvalue. We then 
study the sign of the wave function at each center of a mesh, and calculate
the number of nodal domains by means of the Hoshen-Kopelman 
algorithm\ \cite{Hoshen}. Here, two nearest neighbor meshes with 
the same sign are considered to belong to the same nodal domain. 
The size of the mesh is then changed to a smaller value until the 
convergence of the number of nodal domains is obtained. 
Adopted value of the mesh size is $x_{\rm cl}(E_i)/\max(1.4i,200)$ for
the $i$th eigenstate with energy $E_i$, where 
$x_{\rm cl}(E)$ represents the largest value of the $x$ coordinate for a classical
motion with energy $E$. One should note that the method becomes inaccurate 
for very small values of $k$, where the nodal crossing changes to 
a small avoided crossing due to the nonseparable perturbation.  The smallest positive value 
of $k$ in the present paper is 0.06 which was used in Sec.~\ref{reduction} to 
study the qualitative  behavior of the number of nodal domains by comparing 
with a perturbative argument. 

Contrary to the case of the billiards, the size of the meshed sheet in the present case
is not {\it a priori} determined. 
In fact, since there is no hard wall,  wave functions extend to infinity.
They are, however, rapidly attenuated beyond classically allowed region for given energy. 
Moreover, as shown in Appendix\ \ref{app_proof1}, there appears no new nodal domain 
in the classically forbidden region. Therefore, we adopted the meshed sheet whose 
boundary coincides with that of the classically allowed region. All nodal domains, 
then, have an overlap with the meshed sheet. In case that the  boundary of the 
meshed sheet cuts a nodal domain into several pieces
the number of nodal domains may be overestimated due to the choice of
the meshed sheet. To estimate the 
error caused by this boundary effect, 
we evaluated the difference between the number of nodal domains 
in the adopted meshed sheet and that in the square sheet 
that circumscribes the classically allowed region. 
The estimated difference was 3.1\% for $k=$0.1 and 2.2\% for $k=$0.6.

The Hamiltonian is still  symmetric with respect to the $x$-axis and the 
$y$-axis\cite{Mayer}. We calculated only those wave functions which are symmetric 
with respect to these two axes. 
The diagonalization space was truncated at $n_x+n_y\le 200$, where $n_x$ and $n_y$ 
denote the numbers of oscillator quanta in the  $x$- and $y$-directions. The 
adopted oscillator frequency in the diagonalization was optimized so as to minimize 
the value of ${\rm Tr} H$ in this space for each $k$ value\cite{Aiba}. 

When $k=0.0$, we can obtain more accurate numerical results by adopting 
product of the wave functions for one-dimensional quartic oscillator calculated 
in a similar diagonalization procedure. 
Comparison of the number of nodal domains obtained in the two methods may 
provide a rough estimate for the accuracy of the adopted diagonalization 
procedure, which in the present case is less than 2.4\%.
For the $k=0.0$ results presented below  we adopt those obtained 
by the product wave functions. 

\section{Distribution of the Number of Nodal Domains}
\label{Sec_domain}

\subsection{General consideration on the number of nodal domains}

Before discussing the nodal structure for the present model,
it is useful to study general properties of the number of nodal domains 
which provide a useful classification scheme to be used in later sections. 

The number of nodal domains $N$ for a given nodal line structure of a 
wave function in a two-dimensional area with a boundary $B$ is given by the formula
\begin{equation}
    N(n_b,n_c,m)=\frac12 n_b+n_c+m+1,
    \label{graph}
\end{equation}
under the assumption that more than two nodal lines never cross
at the same crossing point. In Eq.\ (\ref{graph}),  $n_b$ represents the number 
of intersections of nodal lines with the boundary $B$, $n_c$ the number of 
crossing points of  nodal lines, and $m$ the number of islands. Here, 
the term `island' means a cluster of mutually connected nodal lines which 
is linked neither to  $B$ nor to other clusters. 
A similar formula to (\ref{graph}) has been used and proved in 
Ref.\ \cite{Nadirashvili} in a slightly 
different setting of the problem to study the multiplicity of eigenvalues 
for a membrane. In Appendix\ \ref{app_proof2} we give a proof of 
the relation (\ref{graph}) for completeness.

We discuss two cases in which this formula is especially useful. The first is the 
 $m=0$ case, i.e., the case where all nodal lines are linked to the boundary, 
which is typical for separable (and therefore integrable) system. In this case
\begin{equation}
    N=\frac12 n_b+n_c+1. \quad \mbox{(separable, generic)}
    \label{nd_sep}
\end{equation}
This allows us to discuss the dependence of $N$ on the level number in relation to 
the structure of the wave function for $k=0.0$ as shown below. 

The second case is $n_c=0$ which corresponds to the generic wave function of 
nonintegrable systems, where almost all crossings of nodal lines change to 
avoided crossings except for an accidental case. We may first rewrite the 
number of nodal domains in Eq.\ (\ref{graph}) as the sum of two terms:
\begin{equation}
    N=N^{\rm in}+N^b, 
\label{nd_nosep}
\end{equation}
where $N^{\rm in}$ denotes the number of `inner nodal domains', i.e., those which
do not touch the boundary $B$,
 while $N^b$ denotes the remaining part, i.e., the number of ` boundary nodal domains' 
which touch the boundary $B$.
If there is no crossing, $n_c=0$, it is 
easy to see that $N^{\rm in}$ is equal to the number $m$ of islands.
From Eq.\ (\ref{graph}), then, 
we find for $n_c=0$ 
\begin{equation}
    N^{\rm in}=m,\quad N^b=\frac12 n_b+1. 
    \qquad \mbox{(nonintegrable, generic)}
    \label{Nisl_Nb}
\end{equation}
The generic relation Eq.\ (\ref{Nisl_Nb}) for the number of boundary nodal domains 
may provide an estimate on the number of false crossing of nodal lines due to the 
finite mesh size in the numerical calculation.  Comparison of the value of $N^b$ 
with the one obtained from the value of $n_b$ gives a difference of 
 5.2\% for $k=$0.1 \ and 2.5\% for $k=$0.6.
\vspace{-2.5cm}
\begin{figure}[tb]
\begin{center}
\epsfig{file=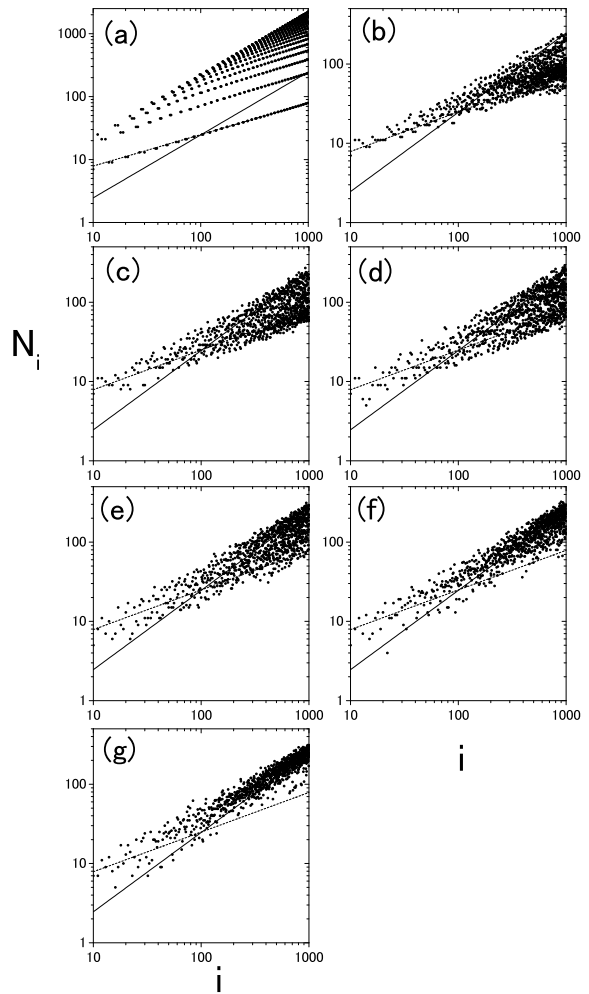,width=10cm}
\end{center}
 \caption{Number of nodal domains $N_i$ vs. the level number $i$ for (a)
$k=0.0$, (b) 0.1, (c) 0.2, (d) 0.3, (e) 0.4, (f) 0.5, and (g) 0.6. 
Solid lines show the function $f(i)$, while the dashed
lines the function $\tilde{f}(i)$. See text for details.
}
\label{Ni_fig}
\end{figure} 
%

\subsection{Numerical results for the distribution of the number of nodal domains}
\label{result}

\begin{figure}
\begin{center}
\epsfig{file=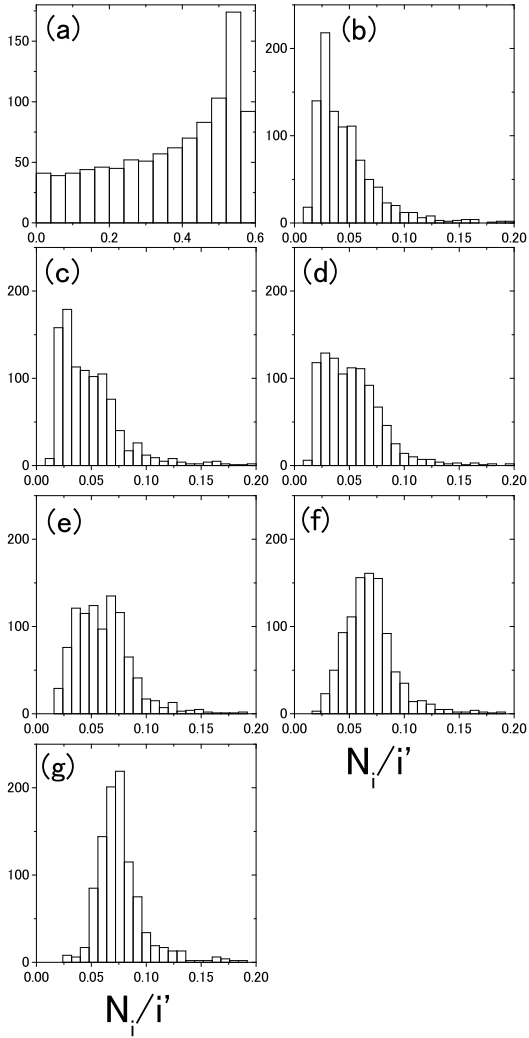,width=10cm}
\end{center}
\vspace{-0.6cm}
 \caption{Histogram of $N_i/i'$ for
(a) $k=0.0$, (b) 0.1, (c) 0.2, (d) 0.3, (e) 0.4, (f) 0.5, and (g) 0.6, where $i'$ stands for
the corrected level number.
}
\label{hist_fig}
\end{figure}

Figure\ \ref{Ni_fig} shows the distribution of the number of nodal domains, 
$N_i$, where the subscript $i$ stands for the level number ordered according 
to the eigenvalue. Note that the quoted level number $i$ is not the 
one of the total system, since we consider only levels which are symmetric 
with respect to the $x$ axis and the $y$ axis. 
In the present model, there are four symmetry classes, and the eigenvalues in the four 
classes are almost equally distributed. 
In the space $n_x+n_y\le 200$ the total number of levels is 20301 and that in the
adopted symmetry class is 5151. Thus, we may assign for each $i$ the corrected 
level number $i'$  approximately given by $i'=ci$ with $c=20301/5151$. 
The solid lines in Fig.\ \ref{Ni_fig} represent the prediction according to the 
percolation-like model which is given by the following function:
\begin{equation}
    f(i)={3\sqrt 3 -5 \over 2}\beta i',
    \label{fi}
\end{equation}
where the value of the coefficient $\beta$ is taken as $2/\pi$ as
calculated for the billiard model \cite{Bogomolny}. 
In the percolation-like model, one first considers a nodal structure of the 
rectangular lattice pattern, and then assumes that an avoided crossing 
of nodal lines occur at every 
lattice point randomly, i.e., either of the positive or negative domains 
is connected at the point with probability 1/2 independent of the other lattice points. 
Figure\ \ref{hist_fig} shows a nodal domain distribution represented as a 
histogram vs. $N_i/i'$. We give  in Table\ \ref{table1} 
the average and the standard deviation of $N_i/i'$. 
\vspace{-1.4cm}

\begin{figure}
\begin{center}
\epsfig{file=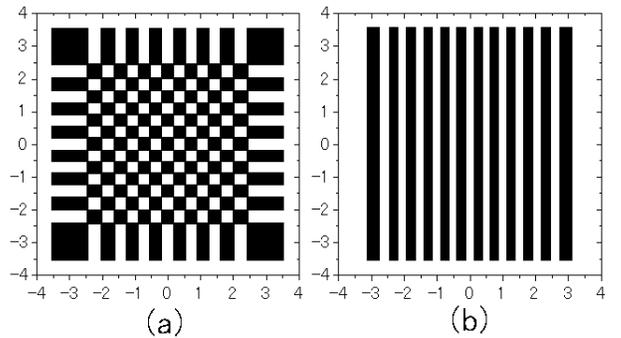,width=8cm}
\end{center}
 \caption{Nodal structures of wave functions at $k=0.0$:
(a) An example of $\psi_{nn}(x,y)$ ($i=98$, $N^{\rm in}_i=169$, $N^b_i=56$)
and (b) an example of $\psi_{n1}(x,y)$ ($i=103$, $N^{\rm in}_i=0$, $N^b_i=25$).
}
\label{patternk00_fig}
\end{figure}

We first discuss the $k=0.0$ case where the system is separable.
In this case the eigenfunction is given by  
$\psi_{mn}(x,y)=\phi_m(x)\phi_n(y)$, where $\phi_m(x)$ represents 
the $m$th wave function of a one-dimensional quartic 
oscillator in the $x$-direction. 
The behavior of the histogram in Fig. \ref{hist_fig} (a) is similar
to the one for the  integrable 
system given in Ref.\ \cite{Blum}, showing an increase and a sharp 
cut-off at some value of $N_i/i'$. By inspecting Fig. \ref{Ni_fig} (a) we 
find that this behavior comes from a number of regular sequences of eigenstates. 
The sequence with the largest $N_i$ values is proportional to $i$, and the 
corresponding wave functions have the form $\psi_{nn}(x,y)$. This may be 
contrasted to the sequence with the smallest $N_i$ values which is 
proportional to $\sqrt{i}$ as shown by the dashed line in Fig.\ \ref{Ni_fig} (a). 
The wave functions in this sequence are of the form $\psi_{n1}(x,y)$ or $\psi_{1n}(x,y)$.
The slopes of other sequences are intermediate between them.
Typical nodal structures corresponding to $\psi_{nn}(x,y)$ and $\psi_{n1}(x,y)$ 
are shown in Fig.\ \ref{patternk00_fig}.
\begin{table}[tb]
\caption{Average and standard deviation of $N_i/i'$ distribution (total)
and those of $N^{\rm in}_i/i'$ distribution (inner, see the text)
for each $k$. Levels $1\le i\le 1000$ are considered.
}
\label{table1}
\begin{center}
\begin{tabular}{lcccr}
&\multicolumn{2}{c}{total}&\multicolumn{2}{c}{inner}\\
$k$&Ave.&S.D.&Ave.&S.D.\\
\tableline
0.0&0.37&0.17&0.29&0.17\\
0.1&0.049&0.035&0.014&0.013\\
0.2&0.051&0.035&0.022&0.016\\
0.3&0.055&0.034&0.031&0.018\\
0.4&0.063&0.032&0.042&0.019\\
0.5&0.071&0.030&0.052&0.018\\
0.6&0.079&0.029&0.060&0.017\\
\end{tabular}
\end{center}
\end{table}
\vspace{-1.0cm}
\begin{figure}
\begin{center}
\epsfig{file=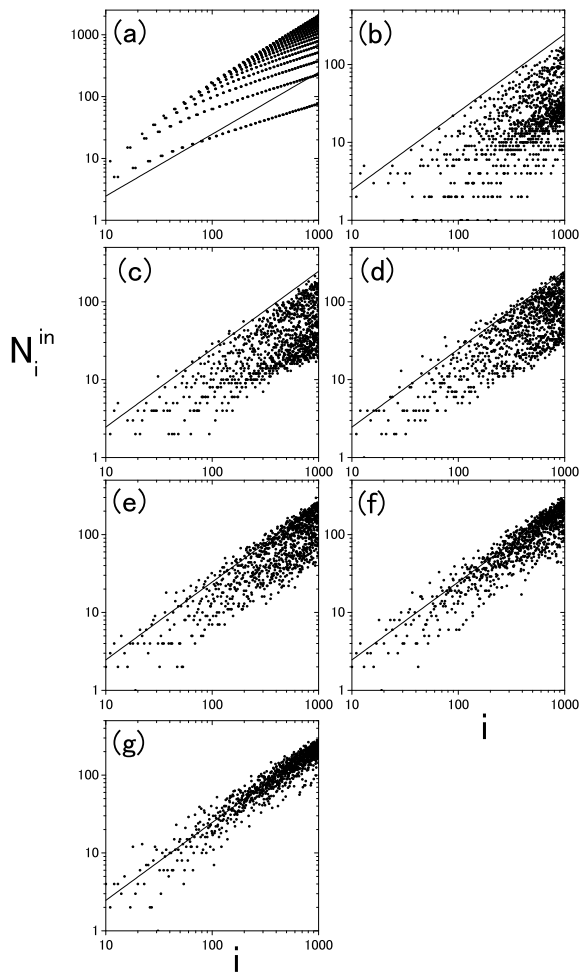,width=10cm}
\end{center}
\vspace{-0.5cm}
 \caption{Number of inner nodal domains $N^{\rm in}_i$ vs. the level number $i$ for (a)
$k=0.0$, (b) 0.1, (c) 0.2, (d) 0.3, (e) 0.4, (f) 0.5, and (g) 0.6. 
Solid lines show the function $f(i)$. 
}
\label{Ni_inner_fig}
\end{figure} 

These behaviors may be understood from  Eq.\ (\ref{nd_sep}).
For eigenfunctions of the form $\psi_{n1}(x,y)$ there is no crossing, $n_c^{(i)}=0$, 
thus $N_i\simeq n_b^{(i)}/2$. 
(Below we explicitly write the  $i$ dependence of $n_b$.)
 Since 
$n_b^{(i)}$ is proportional to  $\sqrt{i}$ \cite{Blum}, the number of nodal
domains $N_i$ in the lowest sequence is also proportional to $\sqrt{i}$.
The dashed line in Fig.\ \ref{Ni_fig} shows the function 
\begin{equation}
   {\tilde f}(i)=d\sqrt{i'},~~d=\frac{2\pi^{1/4}}{\sqrt{3}\Gamma({3 \over 4})}\simeq 1.25,
\label{gi}
\end{equation}
which represents the number of nodal domains for wave functions $\psi_{n1}(x,y)$ 
evaluated in a semiclassical way as given in Appendix\ \ref{app_dergi}.
On the other hand, for eigenfunctions of the type $\psi_{nn}(x,y)$, 
the number of crossing is given by $n_c^{(i)}=(n_b^{(i)})^2/4$, and 
$N_i$ is dominated by $n_c^{(i)}$ if $n_b^{(i)}$ is large enough. 
Accordingly, $N_i$ in the highest sequence is proportional to $i$. 
Note that when the level number $i$ becomes larger, in almost all levels
the contribution of $n_c$ becomes dominant, and the  number of nodal 
domains $N_i$ becomes proportional to $i$ on average.

When the value of $k$ becomes nonzero, e.g., $k=0.1$ in Fig.\ \ref{Ni_fig} (b), 
the number of nodal   
domains is drastically reduced. This is due to the transition from 
the crossing to the avoided crossing of nodal lines. The shape of the 
histogram in Fig.\ \ref{hist_fig} (b) changes, too: The peak at the highest 
value of $N_i/i'$ at $k=0.0$ disappears and the histogram is now 
largely shifted towards low values of $N_i/i'$. 

\vspace{-1.0cm}
\begin{figure}
\begin{center}
\epsfig{file=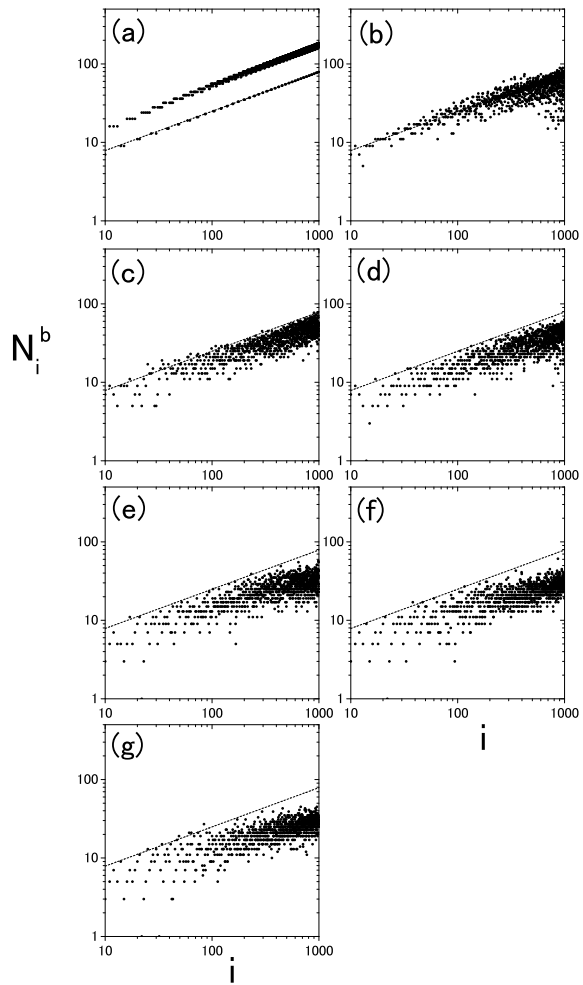,width=10cm}
\end{center}
\vspace{-0.5cm}
 \caption{Number of boundary nodal domains $N^b_i$ vs. the level number $i$ for (a)
$k=0.0$, (b) 0.1, (c) 0.2, (d) 0.3, (e) 0.4, (f) 0.5, and (g) 0.6. 
Dashed lines show the function $\tilde{f}(i)$. 
}
\label{Ni_boundary_fig}
\end{figure} 

\begin{figure}
\begin{center}
\epsfig{file=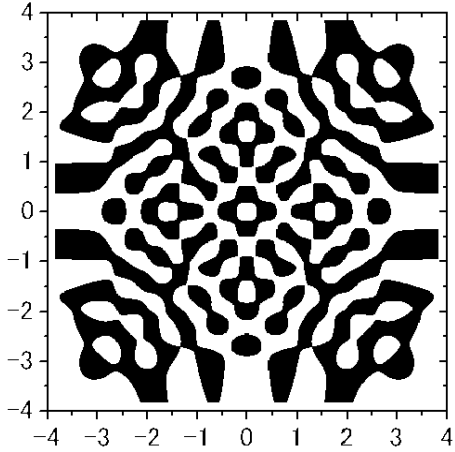,width=6cm}
\end{center}
 \caption{Nodal structure of the 102nd wave function at $k=0.6$,
where $N^{\rm in}_i=30$ and $N^b_i=17$.
}
\label{patternk06_fig}
\end{figure}

To see more details, we  plot $ N^{\rm in}_i$ and $N^b_i$ in 
Figs.\ \ref{Ni_inner_fig} and\ \ref{Ni_boundary_fig}, respectively. 
From these 
figures, we see that at $k=0.1$ the number of inner nodal domains drastically decreases 
below the $f(i)$ line, while the distribution of the number $N^b$ of boundary nodal 
domains tends to accumulate around the ${\tilde f}(i)$ line. These together 
lead to the reduction of the total number of nodal domains compared with 
the separable case at $k=0.0$. The $k$-dependence of $ N^{\rm in}_i$ and that 
of $N^b_i$ are, however, quite different. As $k$ becomes larger, the 
number of boundary nodal domains decreases further and eventually 
becomes rather a small fraction of the total number of domains 
except at low energy (small $i$) region. In contrast, 
the number of inner nodal domains increases again, and in the chaotic 
limit around $k=0.6$ almost aligns to the $f(i)$ line as in the case of 
the billiard system. 
Typical nodal structure for a wave function at $k=0.6$ is shown
in Fig.\ \ref{patternk06_fig}. 

\begin{figure}
\begin{center}
\epsfig{file=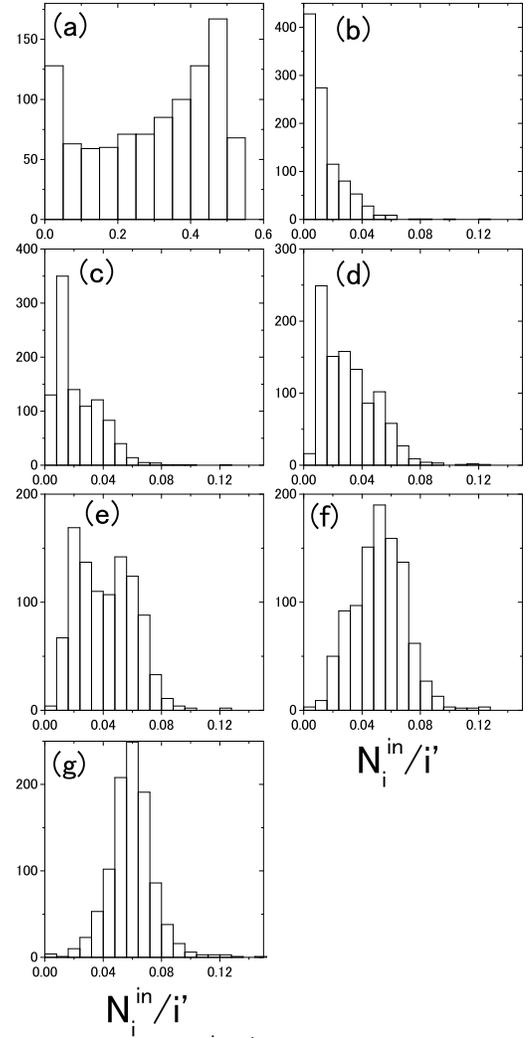,width=10cm}
\end{center}
\vspace{-1.0cm} 
 \caption{Histogram of $N^{\rm in}_i/i'$for
(a) $k=0.0$, (b) 0.1, (c) 0.2, (d) 0.3, (e) 0.4, (f) 0.5, and (g) 0.6, where $i'$ stands for
the corrected level number.
}
\label{hist_inner_fig}
\end{figure}

Let us further study the behavior of the number of inner nodal domains
$ N^{\rm in}_i$. We show the histogram of $N^{\rm in}_i/i'$ in 
Fig.\ \ref{hist_inner_fig}, 
and the average and the standard deviation of $N^{\rm in}_i/i'$ 
in Table\ \ref{table1}. The peak position of the histogram suddenly 
drops almost to zero as $k$ becomes nonzero, which suggests that the 
nonintegrability first acts in such a way to eliminate inner nodal 
domains. This behavior is studied in the next subsection using 
perturbative argument. The peak of $N^{\rm in}_i/i'$ then gradually 
increases with $k$, and finally 
the shape becomes approximate Gaussian at $k=0.6$ centered at finite 
value of  $N^{\rm in}_i/i'$, 
but still is much smaller than the case of $k=0.0$. 
Accordingly, the average of $N^{\rm in}_i/i'$ increases.
The behavior of the histogram of total 
nodal domains at finite $k$ in 
Fig.\ \ref{hist_fig} follows that of the inner nodal domains.

The number of boundary nodal domains $N^b_i$ in 
Fig.\ \ref{Ni_boundary_fig} also shows  interesting features.
At $k=0.0$ there are two regular sequences. The sequence with  larger values
corresponds to the wave functions $\psi_{mn}=\phi_m(x)\phi_n(y)$ 
$(m\ne 1, n\ne 1)$, while the sequence with smaller values to the wave functions 
$\psi_{n1}$ or $\psi_{1n}$. In the latter wave functions 
nodal domains have a one-dimensional structure, while in the former the 
boundary nodal domains are systematically aligned along the boundary.  
Both sequences are proportional 
to the square root of the level number $\sqrt i$.  
In Fig.\ \ref{Ni_boundary_fig}, we show the function 
${\tilde f}(i)$ in Eq.\ (\ref{gi}) by the dashed line.
For small $k$ values, say $k=0.1$, the number of boundary nodal domains 
is distributed around the ${\tilde f}(i)$ line.
\vspace{-1.6cm}
%
\subsection{Reduction of the number of nodal domains in the 
 perturbative regime}
\label{reduction}

When the dynamics changes from integrable to nonintegrable, crossing of nodal lines
changes to avoided crossing, which leads to the reduction of the number of nodal
domains. However, this alone is not sufficient to explain the difference in 
the  reduction at small $k$ value and that at large $k$ value as seen in 
Figs.\ \ref{Ni_inner_fig} and\ \ref{Ni_boundary_fig}. 

\begin{figure}
\begin{center}
\epsfig{file=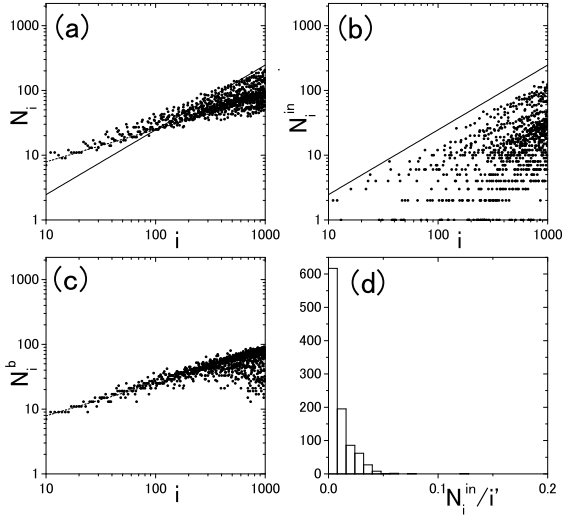,width=10cm}
\end{center}
\vspace{-0.5cm}
 \caption{Number of different kinds of nodal domains at $k=0.06$ is shown against 
 the level number $i$ :
(a) Number of total nodal domains $N_i$, (b) number of inner nodal domains $N^{\rm in}_i$,
and (c) number of boundary nodal domains $N^b_i$. Histogram of $N^{\rm in}_i/i'$ is
shown in (d). Solid lines in (a) and (b) show the function $f(i)$, while dashed line in (a) 
and (c) the function $\tilde{f}(i)$.
}
\label{k006_fig}
\end{figure}

Figure \ref{k006_fig} shows distributions of $N_i$, $N^{\rm in}_i$, 
$N^b_i$, and the histogram of $N^{\rm in}_i/i'$ at $k=0.06$.  We see 
that the number of inner nodal domains is smaller than the case of $k=0.1$,
while the number of boundary nodal domains remains approximately the same.
This further confirms that the number of inner nodal domains becomes smaller as
the value of $k$ decreases as long as the value is nonzero.
As noted above the avoided crossing becomes weaker as the value of $k$ decreases.
Accordingly, the smaller mesh-size is required to avoid the fictitious crossing. 
Indeed, we used the mesh size $x_{\rm cl}(E_i)/\max(3.0i,200)$ for the calculation
at $k=0.06$.  Properties of the avoided crossing (avoidance 
range) have been studied in Ref.\ \cite{Monastra}.
\vspace{-1.0cm}

When the value of $k$ is small, wave functions may well be approximated by 
the 1st order perturbation theory. 
The perturbed wave function can be written as,
\begin{eqnarray}
   \psi'_{mn}(x,y) & = & \psi_{mn}(x,y)+k\sum_{(m',n')\ne (m,n)}C_{m'n'} \psi_{m'n'}(x,y),
   \nonumber \\
   C_{m'n'} & = & \frac{\langle \psi_{m'n'}|x^2y^2|\psi_{mn}\rangle}{E_{m'n'}-E_{mn}}, 
    \label{perturbed_wv}
\end{eqnarray}
where $E_{mn}$ represents an unperturbed energy. Because of  the second term, 
the crossing of nodal lines in the unperturbed wave function $\psi_{mn}$ will 
in generic case turn to an avoided crossing. Whether the positive domains or 
the negative ones are connected and merged into one nodal domain at this avoided crossing 
depends on the 
sign of the second term of (\ref{perturbed_wv}) at the nodal 
crossing point of $\psi_{mn}$. 
Note that if this merging occurs randomly 
independent of the crossing points as in the percolation-like model, one would
not obtain such a huge reduction of the number of inner nodal 
domains as seen in Fig. \ref{k006_fig}.
 This suggests that 
there is a correlation in the sign of the second term among different nodal crossing 
points of $\psi_{mn}$. The sign of the second term is mainly governed by the sign of 
the sum of
a few principal components which have the largest values of $|C_{m'n'}|$. For 
instance, we verified that the probability of the sign of the sum of
two principal components to be equal to that of the total second term at 
randomly chosen points $(x,y)$ is 0.86 for the present model.
Moreover, the difference between the $x$ quantum number of a principal 
component $m'$ and that of the unperturbed wave function $m$, is generally very 
small compared with $m$: $|\delta m|\equiv|m'-m|\ll m$, as long as $m$ is 
large. The same argument holds for the quantum number $n$ in the $y$ direction.

We now evaluate the correlation between signs of a principal component 
with the $x$ quantum number $m'$ at two neighboring crossing points 
along the $x$-direction. The distance $d$ between two neighboring crossing points is 
the typical half wavelength $\frac{1}{2}\lambda_m$ for the quartic oscillator wave function $\phi_m$.
Suppose first that the distance $d$ is shorter than 
the half wavelength $\frac{1}{2}\lambda_{m'}$ of the principal component.
Then, two neighboring crossing points must belong to the same half wavelength 
region in order that signs of the principal component at these two neighboring 
crossing points are same.
Thus, the probability $P_{\rm s}$ that the signs of the principal components at 
two neighboring crossing points to be same can be estimated as
\begin{equation}
P_{\rm s}=
\frac{\lambda_{m'}-\lambda_m}{\lambda_{m'}}=\frac{|\delta m|}{m},
\label{P_s}
\end {equation}
where we used an approximate relation $\lambda_m\propto m^{-1}$.
When the distance $d$ is larger than the
half wavelength $\frac{1}{2}\lambda_{m'}$, the probability $P_{\rm s}$ can also approximately
be given by Eq.\ (\ref{P_s}). 
A superposition of a few principal components also has a wavelength 
close to the unperturbed one $\lambda_m$, leading to the same result. 
Above argument can also be applied to the $y$-direction.
\begin{figure}
\begin{center}
\epsfig{file=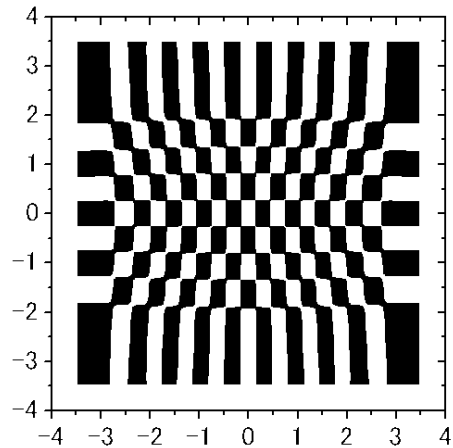,width=6cm}
\end{center}
 \caption{Nodal structure of the 92nd wave function at $k=0.06$,
where $N^{\rm in}_i=0$ and $N^b_i=27$. 
}
\label{patternk006_fig}
\end{figure} 

Eq.\ (\ref{P_s}) shows that the probability $P_{\rm s}$ is much less than unity.
This means that nodal domains in the unperturbed wave function tend to 
be connected along the diagonal when a small perturbation is added, 
which results in the reduction of the
number of inner nodal domains.
Figure\ \ref{patternk006_fig} shows a typical nodal pattern at $k=0.06$.
We see that unperturbed nodal domains, which are slightly deformed from rectangular,
are indeed connected along a diagonal direction. 
(Compare with Fig.\ \ref{patternk00_fig} (a), where $N^{\rm in}_i$ dominates.)
This result supports the above consideration.  

For large $k$ values, the perturbation theory becomes no longer
a good approximation, and the above discussion cannot be applied.

\subsection{Model for the distribution of the number of nodal domains in the chaotic
regime}
\label{model_Ni}

As a reference to the number of nodal domains, we used the function $f(i)$ Eq.\ (\ref{fi})
derived for the chaotic billiard model \cite{Bogomolny}. As previously noted, 
in the percolation-like model, one first considers nodal structure of the rectangular
lattice pattern. In order to determine the size of lattice, the relation $k_x^2=k_y^2=E$ 
was used in Ref. \cite{Bogomolny}, where $k_x$ denotes the wave number of $x$-axis. 
This relation is, however, specific to billiard models. Thus, if a system has a potential
as in the present case, the distribution of the number of nodal domains may differ
from billiard models even in the chaotic regime. Therefore, we would like to present 
a semiclassical model which provides a more general expression for the nodal
domain distribution
incorporating some features of a Hamiltonian with a local potential. 

Let us first summarize the results obtained in Ref.\ \cite{Blum} for 
separable systems, where the Hamiltonian $H$ is 
specified by two action variables $I_1,I_2$.  
According to them, the distribution function of $N_i/i$ in a 
region $E_1\le E_i\le E_2$ (which we call $A$) in the 
$I_1-I_2$ plane can be written  as,
\begin{equation}
   P(\xi)=\frac{1}{N_A}\int_{I_1,I_2\in A}dI_1dI_2\delta\left(\xi-
\frac{\nu(I_1,I_2)}{N(E)}\right),
\label{P_1}
\end{equation}
where $N(E)$ is the semiclassical number of levels
\begin{equation}
   N(E)=\int dI_1dI_2\Theta\left(E-H(I_1,I_2)\right),
\label{NE_1}
\end{equation}
up to energy $E$, $N_A=N(E_2)-N(E_1)$ the number of levels in the region $A$, 
and $\nu(I_1,I_2)=I_1I_2$ is the number of nodal domains 
of the eigenstate specified by $I_1,I_2$. 
Assuming that the Hamiltonian is a homogeneous function of $I_1$ and $I_2$, we
transform the variables from $I_1$ and $I_2$ to the energy $E$ and $r$, where
$r$ represents the length of the path along the arc $\Gamma$ with $E=1$ measured 
from the edge on the $I_2$ axis. 
The action variables are written in terms of the new variables as
$(I_1,I_2)=(g(E)I^{(0)}_1(r),g(E)I^{(0)}_2(r))$, where $I^{(0)}_1(r)$ 
and $I^{(0)}_2(r)$ represent the action on the arc $\Gamma$. 
When the Hamiltonian is a function of the $\ell$th order in $I_1$ and 
$I_2$, $g(E)=E^{1/\ell}$. The Jacobian of the transformation is 
given by the product $J(E)L(r)$,
\begin{eqnarray}
   J(E) & = & 2\,g(E){dg(E)\over dE}, \\
   L(r) & = & \frac12\,\left| \frac{dI^{(0)}_1(r)}{dr}I^{(0)}_2(r)
   -\frac{dI^{(0)}_2(r)}{dr}I^{(0)}_1(r)\right|.
\label{Jacobian}
\end{eqnarray}
With these variables, the number of levels $N(E)$ is given by
\begin{equation}
   N(E)=\left\{ g(E)\right\}^2N(1), \qquad N(1)=\int L(r)dr.
    \label{NE_2}
\end{equation}
By performing the $E$-integration, one obtains the result of Ref.\ \cite{Blum}:
\begin{equation}
   P(\xi)=\frac{1}{N(1)}\int_\Gamma dr\, L(r)\delta
   \left(\xi-\frac{I^{(0)}_1(r)I^{(0)}_2(r)}{N(1)}\right).
 \label{P_3}
\end{equation}

In order to treat the nonintegrable cases, we modify the expression (\ref{P_1}) so 
as to include the following features:
\begin{itemize}
\item[i)]  Hamiltonian is not simply a function of $I_1,I_2$ alone but depends also 
on the angle variables.
\item[ii)] Number of nodal domains for a given energy is reduced on the 
average due to nonintegrability.
\end{itemize}
The first feature implies that each state is not specified by a point $I_1,I_2$ in 
the phase space, but is distributed over an area with fixed $E$ when projected 
on the $I_1-I_2$ plane. We may include this effect by replacing the variable $r$ 
involved in $\nu$ with the smoothed one $r_\omega(r)$;
\begin{equation}
r_\omega(r)\equiv \int_0^{r_{\rm max}}f_\omega(r,r')r'dr',
\label{r_w}
\end{equation}
where, $f_\omega(r,r')$ represents a smoothing function with the width $\omega$, 
and $r_{\rm max}$ the total length of the arc $\Gamma$. The width $\omega$ would 
correspond to the degree of the amplitude mixing of the wave function 
when expanded in the basis of the integrable system. For the integrable system, 
$\omega=0$, and when the system is chaotic, the wave function will be 
distributed over the available phase space, i.e., $\omega\geq r_{\rm max}$.
The second feature ii) may be included by an introduction of a 
reduction factor $G$, i.e., 
\begin{equation}
      \nu(I_1,I_2)\quad\rightarrow\quad I_1I_2G(I_1,I_2)
\end{equation}
with $G=1$ for separable systems. Thus in our model, the distribution 
is given by
\begin{eqnarray}
   P(\xi)&=&\frac{1}{N_A}\int_{E_1}^{E_2}J(E)dE 
   \int_\Gamma dr\,L(r)  \nonumber \\ 
   &\times& \delta\left(\xi-\frac{I^{(0)}_1(r_\omega(r))
                         I^{(0)}_2(r_\omega(r))}{N(1)}\right.  \nonumber \\ 
   &\times& \left.G(I^{(0)}_1(r_\omega(r))g(E),I^{(0)}_2(r_\omega(r))g(E))\,\right).
 \label{P_4}
\end{eqnarray}

To perform the integration, we have to specify $G$.
In the chaotic limit,
the  asymptotic value  $G^{\rm p}$ of $G$ with large $I_1$ and $I_2$ values may be 
obtained from the percolation model of Ref.\ \cite{Bogomolny} as 
\begin{equation}
G(I_1,I_2)\to G^{\rm p}\equiv {3\sqrt{3}-5\over 2} \quad I_1,I_2\to \infty.
\label{G_p}
\end{equation}
On the other hand, $\omega$ becomes large in the chaotic limit. The function $r_\omega$ 
will then become independent of $r$ and is represented by the average, 
i.e., $\bar{r}\equiv r_{\rm max}/2$. By substituting this into 
Eq.\ (\ref{P_4}), we obtain
\begin{equation}
   P(\xi)=\delta\left(\xi-\bar{\xi}\right), \quad
   \bar{\xi}= \frac{I^{(0)}_1(\bar{r})I^{(0)}_2(\bar{r})}{N(1)}G^{\rm p},
 \label{P_5}
\end{equation}
where $\bar{\xi}$ corresponds to the average value of $N_i/i$. 

\begin{figure}
\begin{center}
\epsfig{file=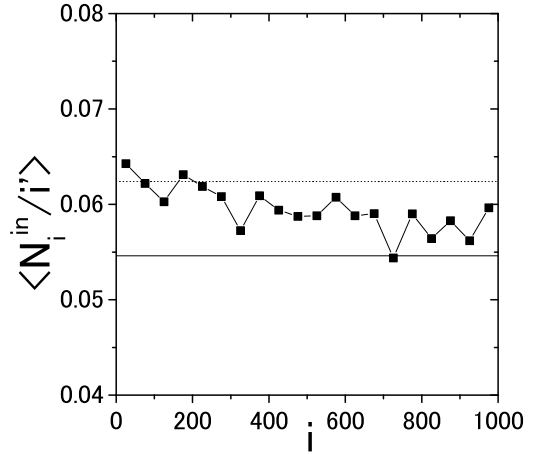,width=8cm}
\end{center}
 \caption{The average $\langle N^{\rm in}_i/i'\rangle$ for $k=0.6$. 
Averaging is performed for bins of $i$ each of which contains 50 levels, 
and the solid squares are placed at the center in each bin. 
The solid line shows the asymptotic value for the quartic
oscillator model and the dashed line for the billiard model.
}
\label{ave_fig}
\end{figure}

We may adopt a concrete example to obtain the value of $\bar{\xi}$. 
Let us consider as an unperturbed model Hamiltonian the following form:
\begin{equation}
H_0\propto I_1^\ell+I_2^\ell.
\label{Hamil_unp}
\end{equation}
The average is calculated as,
\begin{equation}
\left\langle {N_i \over i}\right\rangle=\bar{\xi}=
{2\ell\left(\frac12\right)^\frac{2}{\ell}\Gamma\left(\frac{2}{\ell}\right) \over
\Gamma\left(\frac{1}{\ell}\right)^2}G^{\rm p},
\label{N_ave_def}
\end{equation}
where we used ($\bar{r}=r_{\rm max}/2$) 
\begin{equation}
I_1^{(0)}\left(\bar{r}\right)=I_2^{(0)}\left(\bar{r}\right)=
\left(\frac12\right)^\frac{1}{\ell},
\label{I_k}
\end{equation}
and
\begin{equation}
N(1)=\int_0^1 (1-x^\ell)^\frac{1}{\ell}dx=\frac{1}{\ell}{\Gamma
\left(\frac{1}{\ell}\right)\Gamma\left(\frac{\ell+1}{\ell}\right) \over
\Gamma\left(\frac{\ell+2}{\ell}\right)}. 
\label{N_1}
\end{equation}
Eq.\ (\ref{N_ave_def}) can be applied to the Hamiltonian with or without a potential.
For examples, the average for the billiard model ($\ell=2$) is
\begin{equation}
\left\langle {N_i \over i}\right\rangle=
{3\sqrt{3}-5\over \pi}\simeq 0.062,
\label{ave_bil}
\end{equation}
and that for the quartic oscillator model ($\ell=4/3$) is
\begin{equation}
\left\langle {N_i \over i}\right\rangle=\frac
{\sqrt{2\pi}(3\sqrt{3}-5)}{6\Gamma\left({3\over 4}\right)^2}\simeq 0.055.
\label{ave_quar}
\end{equation}
The difference between the average for the billiard model and that for the quartic
oscillator model is quite small, and cannot be recognized in the log-log plot like 
Fig.\ \ref{Ni_fig}.
Figure\ \ref{ave_fig} compares in detail the result of the numerical calculation and 
the above two asymptotic values. In the numerical calculation we employ 
the number of inner nodal domains $N^{\rm in}_i$
instead of $N_i$ in order to remove the boundary effects. 
The average by numerical calculation still fluctuates between two asymptotic values
and does not seem to converge.
We must investigate higher levels to see if the
actual average converges to the value Eq.\ (\ref{ave_quar}).

In the intermediate region of integrable and chaotic cases, the results for the 
distribution of nodal domain numbers depends on the concrete form of the 
weighting function $f_\omega$ and the reduction factor $G$. 
Here, we consider the specific case where the amplitude mixing is not
complete although large, while the correlation among avoided crossings is lost.
To be more definite, we take the range of $\omega$ as $\frac{1}{2}r_{\rm max}\le \omega<r_{\rm max}$,
and assume that the reduction factor $G$ can be obtained by the percolation model.
In Appendix\ 
\ref{app_ave}, we show that under a reasonable choice of $f_\omega$ we 
obtain the result:
\begin{equation}
     \frac{d}{d\omega}\left\langle\frac{N_i}{i}\right\rangle\ge 0.
\label{der_ave}
\end{equation}
In order to find  which values of $k$ correspond to such a case,
we have to know the relation between $k$, $w$, and $G$, which remains for a future
study. However, the numerical results shown in Sec.\ \ref{Sec_ns} imply that
for $k\geq 0.4$, the assumption of the percolation-like model may be applied. 
Eq.\ (\ref{der_ave}) suggests that the 
average value of $N_i/i$ increases as one approaches the chaotic system, 
which is in accord with the results of Table\ \ref{table1}.
\vspace{-0.5cm}
%

\section{Distribution of the number of nodal intersections}
\label{Sec_isect}

We now consider the behavior of the number of boundary nodal domains, in particular, 
its dependence on the level number  $i$, which can be obtained from that of the number 
of nodal intersections 
with the boundary, $n_b^{(i)}$, according to Eq.\ (\ref{Nisl_Nb}). 

For the case of billiard systems, 
the number of intersections is proportional to $\sqrt i$, even if the dynamics of the 
system is chaotic \cite{Blum}. On the other hand, in Ref.\ \cite{Bies}, the behavior of 
the nodal intersections with the boundary of the classically allowed region was studied 
for systems with soft potentials whose dynamics is chaotic. According to Ref.\ \cite{Bies}, 
the number of intersections per  unit length along the boundary is given by
\begin{equation}
\rho= \gamma|{\rm grad} V|^{1/3},
\label{density}
\end{equation}
where $\gamma=0.171$ and $V$ represents the potential.
\vspace{-0.5cm}
\begin{figure}
\begin{center}
\epsfig{file=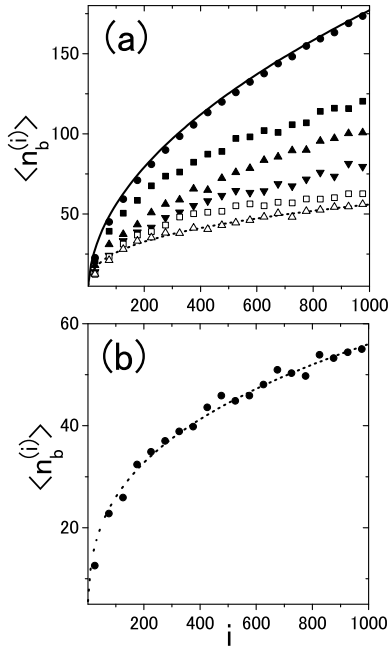,width=8cm}
\end{center}
 \caption{(a) Average of $n_b^{(i)}$ for $k=0.0$, 0.1, 0.2, 0.3, 0.4, and 0.5 from 
top to bottom. Averaging is performed as in the previous figure. Solid and dashed 
lines correspond to $\tau(i)\simeq 5.6i^{1/2}$ and $5.6i^{1/3}$, respectively.
(b) Average of $n_b^{(i)}$ for $k=0.6$. The dashed line corresponds to $5.6i^{1/3}$.
}
\label{nb_fig}
\end{figure}

Now we calculate $n_b^{(i)}$ for the present model, Eq. (\ref{hamiltonian}), by
integrating $\rho$ along the boundary $V(x,y)=E_i$. As the potential is homogeneous,
$n_b^{(i)}$ can be written as,
\begin{equation}
n_b^{(i)}=\gamma \oint_{V=E_i}ds|{\rm grad} V|^{1/3}=\gamma bE_i^{1/2},
\label{nb1}
\end{equation}
with the coefficient $b$ given by
\begin{equation}
b=\oint_{V=1}ds|{\rm grad} V|^{1/3},
\label{b}
\end{equation}
where the integral is performed along the curve $V(x,y)=1$.
By using the semiclassical relation between the energy $E_i$ and the level number $i$
given in Eq.\ (\ref{NE_2}),
we obtain the level number dependence of the number of intersections:
\begin{equation}
n_b^{(i)}=\frac{\gamma b}{N(1)^{1/3}}i'^{1/3}\simeq 3.0i'^{1/3}=4.8i^{1/3},
\label{nb_level}
\end{equation}
where $i'$ represents the corrected level number and  the  numerical value 
$b\simeq 15.26$ evaluated at $k=0.6$ was used.

We note that $n_b^{(i)}$ has a $\sqrt{i}$ dependence at $k=0.0$. This indicates that
the level number dependence of the number of intersections changes from $i^{1/2}$ 
to $i^{1/3}$ as the dynamics changes from integrable to chaotic.
Numerical results shown in Fig.\ \ref{nb_fig} confirm this. Here,
the average value of $n_b^{(i)}$ (over 50 levels) is shown as a function of the level
number $i$. 
At $k=0.0$ $n_b^{(i)}$ follows well the semiclassical result for the number of 
intersections shown by the higher curve in Fig.\ \ref{nb_fig}(a), i.e., 
\begin{equation}
\tau(i)=\frac{8\pi^{5/4}}{5\sqrt{3}\Gamma(\frac{3}{4})^2\Gamma(\frac{5}{4})}\sqrt{i'}
\simeq 2.8\sqrt{i'}=5.6\sqrt{i},
\label{g_2}
\end{equation}
a derivation of which is presented in Appendix\ \ref{app_derg2}.
As $k$ increases, $n_b^{(i)}$ 
gradually decreases. Finally, at $k=0.6$, $n_b^{(i)}$ follows the curve $5.6i^{1/3}$ shown
as the dashed line in Fig.\ \ref{nb_fig} (b), where the fitted value is used as a coefficient. 
The fitted value 5.6 is a little larger than the predicted value 4.8. 
This discrepancy may be due to the curvature of the boundary $V(x,y)=1$, which is not taken
into account in Eq.(\ref{density}). 

\section{Distribution of the Area of the Nodal Domains}
\label{Sec_ns}

Let us now consider the distribution of the area of nodal domains.
The area distribution may provide a measure as to whether the 
percolation-like model can be applied. 
The appropriate length unit to define the area should be the wave 
length\ \cite{Bogomolny} which depends on the energy of the level.
Moreover, in contrast to the billiard problem, the wavelength in the 
present model depend locally on the coordinate 
due to the presence of the potential $V(x,y)$. We define 
the scaled area $s$ of the nodal domain for the level $i$ with energy $E_i$ by
\begin{equation}
s=\mu\int_{\rm n.d.}(E_i-V(x,y))dxdy,
\label{def_area}
\end{equation}
where the integral is performed over each region of the nodal domain. We take 
the scale factor $\mu=1$. One may note that the parameter $\alpha$ is absent in  
the definition (\ref{def_area}) so that it does not directly 
influence the properties of the area distribution. 
The normalized number of nodal domains with area $s$ is defined as\ \cite{Bogomolny},
\begin{equation}
n(s)\equiv\sum_i\frac{Q_i(s)}{i}.
\label{def_ns}
\end{equation}
Here, $Q_i(s)$ represents the number of nodal domains with area $s$ in the 
$i$th wave function.

\begin{figure}
\begin{center}
\epsfig{file=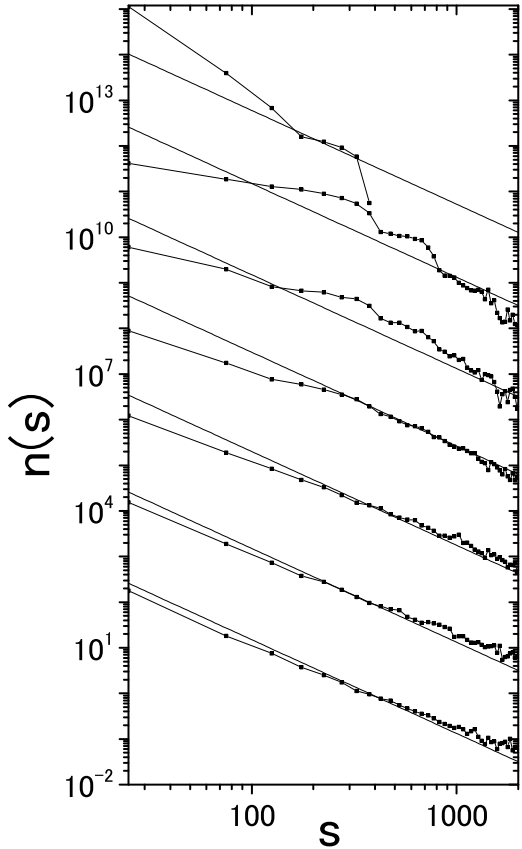,width=8cm}
\end{center}
 \caption{Distribution $n(s)$ for
$k=0.0$, 0.1, 0.2, 0.3, 0.4, 0.5, and 0.6 from top to bottom. The values of $n(s)$
are multiplied by a factor $10^{12}$ for $k=0.0$, $10^{10}$ for $k=0.1$, 
$10^8$ for $k=0.2$, $10^6$ for $k=0.3$, $10^4$ for $k=0.4$, and $10^2$ for $k=0.5$.
Levels $200\le i\le 1000$  are considered. 
Solid lines  show the Fisher line. 
}
\label{ns_fig}
\end{figure}

Figure\ \ref{ns_fig} shows the distribution of $n(s)$. 
The line in the figure represents the power distribution with the Fisher exponent, 
$n^{\rm F}(s)\propto s^{-\tau}$, where $\tau=187/91$, which we call hereafter 
the Fisher line. This represents the characteristic 
distribution at the critical point in the two-dimensional percolation model. 
The figure shows that the distribution at small values of $k$ deviates considerably 
from the Fisher line, especially in the  small area region. 
The deviation gradually diminishes as the value of $k$ increases, and 
at $k=0.6$, the distribution almost coincides with the Fisher line except for 
a small discrepancy in the small $s$ region. We now study if the latter 
discrepancy may imply that the assumption behind the percolation-like 
model does not hold completely even at $k=0.6$.
\vspace{-0.5cm}

We  note that the Fisher line was obtained for an infinite lattice of
mesh points, while 
in our finite system the effect of boundary may influence considerably 
the area distributions. In order to compare with the infinite system and to 
test the applicability of the percolation model, we have to 
exclude the effect of the boundary. To this end, it is appropriate to
employ only the inner nodal domains to obtain the area distribution.

\begin{figure}
\begin{center}
\epsfig{file=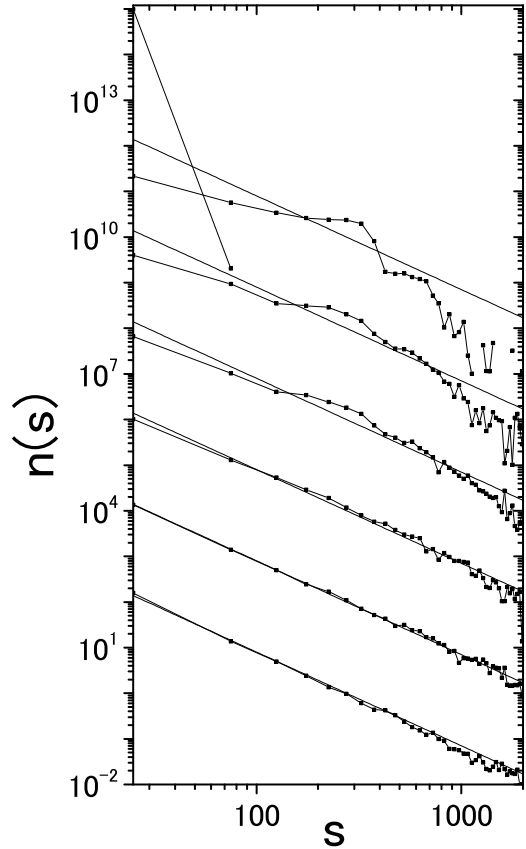,width=8cm}
\end{center}
 \caption{Same as Fig.\ \ref{ns_fig} but for the inner nodal domains.
The Fisher line corresponding to $k=0.0$ is omitted.
}
\label{ns_inner_fig}
\end{figure}

Figure\ \ref{ns_inner_fig} shows the area distribution $n(s)$ for inner nodal domains.
Large deviation from the Fisher line at small $k$ values reflects the fact that the avoided
crossings are correlated and the nodal structure does not belong to the universality class
of the percolation model. On the other hand, when $k\ge 0.4$, the agreement is very good.
Thus we may conclude that the small deviation seen in Fig.\ \ref{ns_fig} at $k\ge0.4$ is due to
the boundary effect.
\vspace{-0.5cm}

Remember that the distribution of the number of nodal domains 
in Fig.\ \ref{hist_fig} 
and the number of inner nodal domains in Fig.\ \ref{hist_inner_fig}
still show a gradual change after $k=0.4$. 
The area distribution is contrasted to this behavior. 
The agreement with the Fisher line already at $k\ge0.4$ may imply that the independence 
and the randomness of the avoided crossings is practically already realized before
the system becomes completely chaotic at $k=0.6$ as given by the classical phase space structure 
or by the level statistics. 

\section{Summary}
\label{Sec_summary}

We investigated the transition from integrable to chaotic dynamics from 
the point of view of the nodal structure in the  wave functions by
employing a two dimensional quartic oscillator. 
The distribution of the number of nodal domains, the number of intersections
of nodal lines with the boundary of the classically allowed region, and the area distribution
of nodal domains were studied as a function of the parameter $k$ which controls the 
nonintegrability of the system.

The number of nodal domains is drastically reduced as the dynamics of the system 
changes from integrable ($k=0.0$) to nonintegrable ($k$ nonzero), 
and then gradually increases as the system becomes chaotic ($k\sim 0.6$). Separation 
into `inner' and 
`boundary' nodal domains shows that the above dependence on $k$ mainly comes from the 
behavior of the former.  Perturbative argument suggests that a finite (but small) $k$ gives 
rise to  correlated avoided crossings of nodal lines, leading to a drastic reduction of 
inner nodal domains. Their number turns to increase again as $k$ becomes larger, which 
may be related to the loss of the correlations in avoided crossings which underlies the 
percolation model. The `boundary' nodal domains show, in contrast, a milder dependence on 
$k$, and its  significance in the total number of nodal domains becomes small at 
large $k$ values. A semiclassical model which incorporates the degree of amplitude 
mixing as well as properties of avoided crossings has been proposed to study the 
number of nodal domains.

We studied the distribution of the number of intersections with the boundary of 
the classically allowed region. We found that the average number shows a different 
dependence on the level number
as the dynamics changes from integrable to chaotic. It is interesting to  find such 
a characteristic connection to the dynamics in the structure of the wave function 
at the boundary, in view of the  rather mild $k$ dependence found in the number of 
boundary nodal domains.

We studied also the distribution of the nodal domain areas which shows
a scaling behavior in the percolation-like model\cite{Bogomolny}. In the 
present model, too, the area distribution shows a scaling with the same exponent 
for large $k$ values. A small deviation has been shown to come from the boundary effect. 
The scaling behavior seems to be complete before the dynamics reaches the
chaotic limit, however. 

This raises the question as to how the various signatures on the chaotic properties 
which appear in the level statistics or in the wave functions are interrelated. 
Further studies on the typical values of $k$ which 
characterize the onset or the completion of various structures in the wave functions 
would be interesting. They include, for instance, 
the signals studied in the nodal domain distribution, amplitude 
mixing (Porter-Thomas distribution), loss of correlations in the avoided crossings, 
scaling in the area distribution, etc. These are left for the  future investigation. 
In this context, it is valuable to mention that even for the superposition of random waves,
there is a weak long range correlation among the avoided crossings as discussed
in Ref.\ \cite{Foltin}. 

\acknowledgments

The authors acknowledge Prof. O. Bohigas for suggesting to the authors the 
study of nodal domain distributions in the anharmonic oscillator model. They  
thank Prof. H. Shudo and Prof. A. Tanaka for a discussion and Prof. S. Mizutori 
for a valuable comment. They also acknowledge Prof. N. Nadirashvili who kindly 
informed us of his previous works in Ref.\ \cite{Nadirashvili}.
\appendix
\section{No nodal domain in the classically forbidden region}
\label{app_proof1}

We prove that there is no nodal domain which is embedded entirely in a classically 
forbidden region.

Assume that there is such a nodal domain whose region is $D$ for an eigenfunction 
$\Psi$ (taken real, for simplicity).
The eigenfunction $\Psi$ satisfies the following Schr\"odinger equation;
\begin{equation}
\Delta\Psi=2(V-E)\Psi,
\label{s_eq}
\end{equation}
where $E$ represents the eigenvalue and $V$ the potential.
Multiply  $\Psi$ on both sides in Eq.\ (\ref{s_eq}) and integrate them over 
the region $D$. The right hand side becomes positive,
\begin{equation}
2\int_D(V-E)\Psi^2 dv>0,
\label{rhd}
\end{equation}
because in the classically forbidden region the inequality $V-E>0$ always holds.
On the other hand, since the value of $\Psi$ along the boundary of the region $D$
is zero, the left hand side after partial integration becomes 
\begin{equation}
\int_D\Psi \Delta\Psi dv=-\int_D (\nabla\Psi)^2dv<0,
\label{lhd}
\end{equation}
which is in contradiction to Eq.\ (\ref{rhd}).
Therefore there is no nodal domain in the classically forbidden region. 

\section{Number of nodal domains for a given diagram of nodal lines}
\label{app_proof2}

We derive the relation (\ref{graph}) for the number 
of nodal domains in a two-dimensional area enclosed by a boundary $B$. 
See also Ref.\ \cite{Nadirashvili}  for a slightly different form of the relation 
for the number of nodal domains expressed in terms of the number of 
nodal lines, etc. 
The nodal lines and the boundary represent a kind of a diagram in a two-dimensional 
plane which is constructed with a number of vertices (nodal crossings including 
nodal contacts at the boundary) that are connected with edges, i.e., the line segments 
of nodal lines or those of the boundary. 
We first consider a general diagram made of nodal lines and then restrict it to the 
special case treated in the text. 

Let us define the degree $k$($\ge 3$) of a vertex as the number of lines which are connected 
at the vertex. We define also `island' (in $B$) as the cluster of nodal lines which are 
linked neither to $B$ nor to the other clusters of nodal lines. Simplest island is 
a `bubble', i.e., a closed line without a vertex. For instance, a bubble within a bubble 
will make two islands. 

We now consider the number $N$ of nodal domains for 
a diagram of nodal lines in $B$ with $m$ islands, where 
the total number of vertices of degree $k$($\ge 3$) is $n_k$. Since there are $m+1$ 
clusters of nodal lines disconnected from each other, we assign them an index 
$j=0,1,\cdots,m$, where $j=0$ denotes the cluster which involves the boundary $B$, while 
$j=1,\cdots,m$ denote islands. For each cluster $j$ we assign the number of 
nodal domains $N_j$, the total number of edges $e_j$, and that of vertices $v_j$
which is given by the sum of $n_k(j)$, the number of vertices with degree $k$ in 
the $j$-th cluster. 
If we neglect the case of `bubbles' for a moment, we can 
use Euler's formula for each cluster $j$ to obtain
\begin{equation}
   N_j+1 =e_j-v_j+2, \qquad (j=0,1,2,\ldots,m)
   \label{euler}
\end{equation}
where we added unity on the left hand side, since $N_j$ counts domains only 
inside the boundary. By using the graphical relation 
\begin{equation}
    e_j=\frac12\sum_{k\ge 3}k\,n_k(j),
\end{equation}
the relation (\ref{euler}) is transformed to 
\begin{equation}
    N_j=\sum_{k\ge 3}\left(\frac12 k-1\right)n_k(j)+1.
    \label{jthdomain}
\end{equation}
Equation (\ref{jthdomain}) is seen to hold also for a `bubble', where 
$N=1$ and all $n_k$'s are zero. By summing up over $j$ we 
obtain the relation
\begin{equation}
    N=\sum_{j=0}^mN_j=\sum_{k\ge 3}\left(\frac12 k-1\right)n_k+m+1,
    \label{Ntotal}
\end{equation}
where we used $n_k=\sum_jn_k(j)$.

In the special case treated in Sec.\ \ref{Sec_domain}, we assumed that there are no 
accidental crossing of more than two nodal lines at a point inside $B$, 
and also that no accidental contact of more than one nodal line on $B$. 
In this case, the vertex inside $B$ has a degree 4, and the vertex on $B$ has a 
degree 3. Thus the number of nodal domains $N(b,c,m)$ 
for a diagram with $c(=n_4)$ crossings, $b(=n_3)$ points of contact on $B$, and $m$ islands
is given by 
\begin{equation}
    N(b,c,m)=\frac12 b+c+m+1,
\end{equation}
which is  the relation (\ref{graph}).

\section{Derivation of some relations}
\label{app_semi}

We give brief derivations for Eqs.\ (\ref{gi}) and (\ref{g_2}). 
See Sec.\ \ref{model_Ni} for notations.

\subsection{A derivation of Eq.\ (\ref{gi})}
\label{app_dergi}

For the wave function $\psi_{n1}(x,y)$ which has corresponding actions, $I_1=I$, $I_2=0$,
the semiclassical number of nodal domains $N_i$ can be written  as,
\begin{equation}
N_i=I=g(E),
\label{Ni_semi}
\end{equation}
while the semiclassical level number is given by Eq.\ (\ref{NE_2}).
Then, the coefficient $d$ in Eq.\ (\ref{gi}) is given by
\begin{equation}
d=1/\sqrt{N(1)},
\label{coef_d}
\end{equation}
which leads to Eq.\ (\ref{gi}) for the quartic oscillator model.

\subsection{A derivation of Eq.\ (\ref{g_2})}
\label{app_derg2}

Analogous to Eq.\ (\ref{P_1}), the distribution function of $n_b^{(i)}/\sqrt{i}$
can be written as,
\begin{equation}
   P(\eta)=\frac{1}{N_A}\int_{I_1,I_2\in A}dI_1dI_2\delta\left(\eta-
\frac{\tau(I_1,I_2)}{\sqrt{N(E)}}\right),
\label{P_isect}
\end{equation}
where $\tau(I_1,I_2)=2(I_1+I_2)$ is the number of nodal intersections.
By the same variable transformation as in Sec.\ \ref{model_Ni} and by
performing the integration over the variable $E$, the average of $\eta$ can be
written as,
\begin{equation}
\langle \eta \rangle=\frac{2}{N(1)^{3/2}}\int_\Gamma drL(r)(
I_1^{(0)}(r)+I_2^{(0)}(r)).
\label{ave_eta}
\end{equation}
Inserting Eq.\ (\ref{Jacobian}) and transforming the variable $r$ to $I_1^{(0)} (=x)$,
the integral in Eq.\ (\ref{ave_eta}) can be expressed as,
\begin{eqnarray}
\int_0^1 dx x(1-x^\ell)^{\frac{1}{\ell}}
+\int_0^1 dx(1-x^\ell)^{\frac{2}{\ell}} \nonumber \\
=\frac{\Gamma\left(\frac{2}{\ell}\right)
\Gamma\left(\frac{\ell+1}{\ell}\right)}
{\ell\Gamma\left(\frac{\ell+3}{\ell}\right)}
+\frac{\Gamma\left(\frac{1}{\ell}\right)
\Gamma\left(\frac{\ell+2}{\ell}\right)}
{\ell\Gamma\left(\frac{\ell+3}{\ell}\right)},
\label{int2}
\end{eqnarray}
for the Hamiltonian with the form Eq.\ (\ref{Hamil_unp}).
Thus, for the quartic oscillator model ($\ell=4/3$), Eq.\ (\ref{ave_eta})
gives the coefficient in Eq.\ (\ref{g_2}).  

\section{Example for the calculation of the average of nodal domain numbers}
\label{app_ave}

We here show the increase of average of $N_i/i$  after $\omega\ge\frac{1}{2}r_{\rm max}$ 
by using Eq.\ (\ref{P_4}) with the assumption that the reduction factor $G$ can be
obtained from the percolation model.
The reduction factor $G(I_1,I_2)$ may still have a dependence on $I_1$ and $I_2$, however, when
the energy is not so high.

As for the smoothing function, we adopt the following form:
\begin{equation}
f_\omega(r,r')={\theta(\omega-|r-r'|)\over \min(r+\omega,r_{\rm max})-\max(r-\omega,0)}.
\label{f_w}
\end{equation}

The average of $N_i/i$ can be written as,
\begin{eqnarray}
\left\langle{N_i\over i}\right\rangle={1\over N_A}\int_{E_1}^{E_2}J(E)dE 
\int_\Gamma drL(r) \nonumber \\
\times {I^{(0)}_1(r_\omega)I^{(0)}_2(r_\omega)\over N(1)} G(I^{(0)}_1(r_\omega)g,I^{(0)}_1(r_\omega)g),
 \label{xi_ave}
\end{eqnarray}
where $r_\omega=r_\omega(r)$ and $g=g(E)$.
We differentiate Eq.\ (\ref{xi_ave}) with respect to $\omega$,
\begin{eqnarray}
\frac{d}{d\omega}\left\langle{N_i\over i}\right\rangle =
{1\over N_A}\int_{E_1}^{E_2}J(E)dE 
\int_\Gamma drL(r) \nonumber \\
\times{r_\omega' \over N(1)}C(E,r) \label{diff_ave}\\
C(E,r)=(I_1^{(0)\prime}(r_\omega)I_2^{(0)}(r_\omega)+I_1^{(0)}(r_\omega)I_2^{(0)\prime}(r_\omega))G
\nonumber \\
+I_1^{(0)}(r_\omega)I_2^{(0)}(r_\omega)g({\partial G \over \partial a}I_1^{(0)\prime}+{\partial 
G \over \partial b}I_2^{(0)\prime}),
\label{C_Er}
\end{eqnarray}
where $r_\omega'=dr_\omega(r)/d\omega$, 
$I_1^{(0)\prime}(r)=dI_1^{(0)}(r)/dr$, and $\partial G/\partial a=
\partial G(a,b)/\partial a$ etc.

When $\omega>\frac{1}{2}r_{\rm max}$, the function $r_\omega'$ has a following form;
\begin{equation}
r_\omega'(r)=\left\{
\begin{array}{rl}
{1\over 2},& \quad (r\le r_{\rm max}-\omega)\\
0,& \quad (r_{\rm max}-\omega< r\le \omega)\\
-{1\over 2},& \quad (r>\omega).
\end{array}\right.
\label{der_rw}
\end{equation}
Accordingly, the integration range of $r$ in Eq.\ (\ref{diff_ave}) is restricted to 
$(0,r_{\rm max}-\omega)$ and $(\omega, r_{\rm max})$.
By using the relation $G(a,b)=G(b,a)$ and approximate relations
$I^{(0)}_1(r)=I^{(0)}_2(r_{\rm max}-r)$ and $I^{(0)\prime}_1(r)=-I^{(0)\prime}_2(r_{\rm max}-r)$,
the latter integration range can be transformed to the former as,
\begin{eqnarray}
\frac{d}{d\omega}\left\langle{N_i\over i}\right\rangle=
{1\over N_AN(1)}\int_{E_1}^{E_2}J(E)dE \nonumber \\
\times \int_0^{r_{\rm max}-\omega} drL(r)C(E,r). 
\label{diff_ave2}
\end{eqnarray}
Above approximate relations are justified by an approximate symmetry between
the variables $I_1$ and $I_2$, because the symmetry breaking in the 
Hamiltonian\ (\ref{hamiltonian}) is small.
The behavior of the average is, thus, determined by the function $C(E,r)$.
The function $C(E,r)$ can be written as,
\begin{eqnarray}
C(E,r)=I^{(0)\prime}_1I^{(0)}_2\Biggl\{\left(1+{I^{(0)}_1\over I^{(0)}_2}{dI^{(0)}_2\over
dI^{(0)}_1}\right)G \nonumber \\
+I^{(0)}_1g\left({\partial G\over \partial a}+
{dI^{(0)}_2\over dI^{(0)}_1}{\partial G\over \partial b}\right)\Biggr\}.
\label{C_Er2}
\end{eqnarray}
Since the range of $r_\omega$ is $\frac{1}{4}r_{\rm max}\le r_\omega\le \frac{1}{2}r_{\rm max}$
for $\omega\ge \frac{1}{2}r_{\rm max}$,
where the boundary effect is negligible,
the value of the reduction factor $G$ will be close to the asymptotic one.
Thus, the derivative of $G$ is near zero;
\begin{equation}
{\partial \over \partial a}G(a,b)={\partial \over \partial b}G(a,b)\simeq 0.
\label{der_G_0}
\end{equation}
Accordingly, the second term in Eq.\ (\ref{C_Er2}) is neglected.
We also assume the Hamiltonian for integrable case has the form, Eq.\ (\ref{Hamil_unp}).
The function $C(E,r)$ ,then, can be approximately written as,
\begin{equation}
C(E,r)\simeq I^{(0)\prime}_1I^{(0)}_2\left(1-\left({I^{(0)}_1\over I^{(0)}_2}\right)^\ell\right)G.
\label{appro_C_Er}
\end{equation}

Since in the considered range of $r_\omega$, the relation $I^{(0)}_1\le I^{(0)}_2$ always holds,
the derivative of average with respect to $\omega$ is
positive,
\begin{equation}
\frac{d}{d\omega}\left\langle{N_i\over i}\right\rangle\ge 0.
\label{der_ni_posi}
\end{equation}
This result shows that the average of $N_i/i$ increases as the smoothing width $\omega$, 
as long as $\omega\ge \frac{1}{2}r_{\rm max}$. 
The increase of the average stops when the condition $I^{(0)}_1=I^{(0)}_2$ always holds,
which means $r_\omega=\frac{1}{2}r_{\rm max}$, namely the chaotic limit.



\begin{references}
\bibitem{Berry} M.V. Berry, J. Phys. {\bf A 10}, 2083(1977).
\bibitem{McDonald} S.W. McDonald and A N. Kaufman, Phys. Rev. Lett. {\bf 42}, 1189 (1979).
\bibitem{Brody} T.A. Brody, et al.,  Rev. Mod. Phys. {\bf 53}, 385 (1981).
\bibitem{Blum}G. Blum, S. Gnutzmann, and U. Smilansky, Phys. Rev. Lett.
{\bf 88}, 114101 (2002).
\bibitem{Bogomolny}E. Bogomolny and C. Schmit, Phys. Rev. Lett. 
{\bf 88}, 114102 (2002).
\bibitem{Savytskyy} N. Savytskyy, O. Hul, and L. Sirko, Phys. Rev. E {\bf 70}, 056209
(2004).
\bibitem{Mayer} See, for instance, H.-D. Mayer, J. Chem. Phys. {\bf 84}, 3147 (1986).
\bibitem{levels} Th. Zimmermann, H.-D. Meyer, H. Koppel, and L. S. Cederbaum, 
Phys. Rev. A {\bf 33}, 4334 (1986).
\bibitem{Hoshen}J. Hoshen and R. Kopelman, Phys. Rev. {\bf B 14}, 3438 (1976).
\bibitem{Aiba} H. Aiba and T. Suzuki, Phys. Rev. {\bf E 63}, 026207 (2001).
\bibitem{Nadirashvili}T. Hoffmann-Ostenhof, P. Michor, and N. Nadirashvili, 
         Geom. Funct. Anal. {\bf 9}, 1169 (1999);  
        N. Nadirashvili, Math. USSR Sbornik {\bf 61}, 225 (1988).
\bibitem{Monastra} A. G. Monastra, U. Smilansky, and S. Gnutzmann, 
J. Phys. A {\bf 36}, 1845 (2003).
\bibitem{Bies} W. E. Bies and E. J. Heller, J. Phys. A {\bf 35}, 5673 (2002).
\bibitem{Foltin} G. Foltin, S. Gnutzmann, and U. Smilansky, J. Phys. {\bf A 37}, 11363 (2004).
\end{references}
\end{document}